# TEMPORAL ASPECTS OF SOLAR FLARES AND NEUPERT EFFECT DURING OCTOBER-NOVEMBER 2003

Navin Chandra Joshi[1], Neeraj Singh Bankoti, Seema Pande, Bimal Pande, Kavita Pandey
*Department of Physics, DSB Campus, Kumaun University, Naini Tal - 263 001, Uttarakhand, India*
[1]njoshi98@gmail.com

**ABSTRACT:**
A statistical analysis of dataset of Hα, soft X-ray (SXR) and Hard X-ray (HXR) flares comprising almost 318, 401 and 1920 single events respectively that occurred during the period of October-November 2003 is presented. On the basis of this dataset, statistics on temporal properties e.g. duration, rise and decay times of Hα, SXR and HXR flare with regard to the flare classes is presented. The duration, rise and decay time for Hα flares increases with increasing importance and brightness classes. The increase is more pronounced for the duration and decay time. The same relation is valid with regard to the brightness class. On the other hand for SXR flares, the duration, rise and decay also increase with the flare class being more pronounced for the duration and rise time. We have also analysed the event asymmetry (EA) of Hα, SXR and HXR flares for the same period. The EA indices are predominantly positive for Hα but for SXR and HXR this shows symmetrical distribution on both sides of zero asymmetry. Our study also presents the result of the Neupert effect (NE) in solar flares for the same period. We investigate the time difference (Δt) between the maximum of SXR emission and the end of the HXR emission for many flares during the period of our investigation. Our study shows that distribution of the time difference reveals a peak at $\Delta t = -1$. Out of 236 events, only 48 (20.34 %) events satisfy the NE whereas there is also a significant fraction of flares 62 (26.27 %) which shows strong deviation form $\Delta t = 0$. The period of investigation is focused on the descending phase of solar cycle 23. Study of this period is significant because in this decay phase activity exhibits a sudden increase.

## 1. INTRODUCTION

Statistical investigations on temporal aspects of solar flares observed in various soft X-ray (SXR) wavelengths have been carried out by various authors (Culhane and Phillips 1970; Drake 1971; Thomas and Teske 1971; Datlowe, Hudson and Peterson 1974; Pearce and Harrison 1988; Veronig et al, 2002a). Temporal aspects of solar flares observed in optical (Hα) have also been studied by many authors (Newton and Barton 1937; Waldmeier 1938, 1948; Giovanelli 1948; Ellison 1949; Warwick 1954; Dodson, Hedeman and McMath 1956; Waldmeier and Bachmann 1959; Smith 1962; Reid 1968; Ružičková-Topolová 1974; Wilson 1983, 1987; Antalová 1985; Barlas and Altas 1992; Temmer et al 2001). Temmer et al. (2001) statistically analysed a large dataset of Hα flares during the period January 1975 to December 1999. Veronig et al. (2002a) presented a statistical analysis of SXR flare during the period 1976-2000. Observations show that the shape of the rising phase of the SXR light curve closely resembles the time integral of the microwave or hard X-ray light (HXR) curve. It has been argued that this observation is evident for a causal relation between the non thermal HXR emission and the thermal SXR emission. This is known as Neupert effect (NE) (Neupert 1968; Hudson 1991; Dennis and Zabro 1993). Much work was devoted to the study of this subject beginning with Neupert (1969). Dennis and Zabro (1993) Studied NE using the actual SXR and HXR light curves. Recently Veronig et al (2002b) investigated the NE in solar flares by utilizing statistical properties of solar flares observed simultaneously in SXR and HXR.

In addition to flare activity other activity parameters have also been analysed by various authors during 2003 October-November (Zhang et al., 2003; Gopalseamy et al., 2005; Schmider et al., 2006). Gopalswamy et al (2005) presented a statistical analysis of the Coronal Mass Ejections (CME) and compared them with those of the general population of CME observed during solar cycle (SC) 2003. Magnetic configuration of three super active regions NOAA 10484, 10486 and 10488 from 2003 October 18 to November 4 have been analysed (Zhang et al., 2003).

In the present paper, an attempt is made to study the temporal aspects of flares (Hα, SXR, and HXR) and NE along with the statistical properties of flares during the October-November 2003. The paper is organized as follows. In Section 2 we have described the Hα, SXR and HXR dataset used in the analysis. Section 3 describes the applied statistical methods. Different statistical investigations of temporal properties of Hα flares (Section 4.1), SXR flares (Section 4.2) and HXR flares (Section 4.3), event asymmetry (EA) (Section 4.4) and NE (Section 4.5) are presented. The final section (5) deals with discussions and conclusions.

## 2. DATA

For the present analysis we make use of the Hα and SXR flare data listed in the National Geophysical Data Center (NGDC) from 01 October 2003 to 30 November 2003. During this period of 61 days (two months) 401 Hα flares and 323 SXR flares are reported. For some Hα events the peak time is not given. Excluding these events, we get a total of 318 Hα events for analysis. The URL address to obtain these data is
*ftp://ftp.ngdc.noaa.gov/STP/SOLAR_DATA/SOLAR_FLARES*. Table 1 lists the number of Hα flares corresponding to different importance classes (S=subflares, or 1, 2, 3 or 4 for successively larger flares) denoting the size of the flare. Table 2, the number of flares are subdivided into three brightness classes (f=faint, n=normal, b=bright) corresponds to a subjective estimate of the intensity of emission. In Table 3, we list the number of flares belonging to different groups of importance and brightness classes resulting in nine subclasses. From Table 3, it is clear that subflares and importance 1flares have a strong tendency to be faint and importance > 1 flares to be of bright.

**Table 1.** The number of flare events for the different Hα flare classes (S, 1, 2, 3, and 4) and the corresponding percentage values. T denotes the total number of flares occurring in the period (Oct-Nov 2003) chosen for study.

| Importance | No. of events | % |
|---|---|---|
| S | 267 | 83.96 |
| 1 | 34 | 10.69 |
| 2 | 12 | 3.77 |
| 3 | 3 | 0.94 |
| 4 | 2 | 0.63 |
| T | 318 | 100.00 |

**Table 2.** The number of flare events for the different Hα flare classes (f, b, n) and the corresponding percentage values. T denotes the total number of flares occurring in the selected period (Oct-Nov 2003).

| Brightness | No. of events | % |
|---|---|---|
| f | 285 | 89.62 |
| b | 23 | 7.23 |
| n | 10 | 3.14 |
| T | 318 | 100.00 |

**Table 3.** The number of flare events for the different importance (S, 1, >1) and brightness classes, given in absolute values and percentage for different respective classes.

| | S | | Imp. 1 | | Imp. >1 | |
|---|---|---|---|---|---|---|
| f | 260 | (97.38 %) | 21 | (61.76 %) | 4 | (23.53 %) |
| n | 7 | (2.62 %) | 12 | (35.29 %) | 4 | (23.53 %) |
| b | 0 | (0.00 %) | 1 | (2.94 %) | 9 | (52.94 %) |
| T | 267 | (100.00 %) | 34 | (100.00 %) | 17 | (100.00 %) |

It can be clearly seen in Table 1 that the percentage of flares of importance classes 2, 3, and 4 is very small compared to subflares and importance class 1 flares. Therefore, for the purpose of analysis these importance classes are combined into one group and denoted as importance >1. Table 4 lists the number of SXR flares reported for the concerned period, subdivided into different SXR flare classes. From Table 4, it can be seen that most of the flares belong to class C. Larger flares, i.e. M and X are less in numbers.

**Table 4.** The numbers of flare events for the different SXR flare classes (B, C, M, X) and the corresponding percentage values. T denotes the total number of flares occurring in the period (Oct-Nov 2003) selected for the study.

| Class | No. of events | % |
|---|---|---|
| B | 106 | 26.43 |
| C | 223 | 55.61 |
| M | 61 | 15.21 |
| X | 11 | 2.74 |
| T | 401 | 100.00 |

To study NE we have analysed time difference between the maximum of SXR emission and the end of HXR emission. We utilize *GOES (Geostationary Operational Environment Satellite)* SXR data in the 0.1-0.8 nm wavelength band as listed in NGDC. The URL of the web site is *ftp://ftp.ngdc.noaa.gov/STP/SOLAR_DATA/SOLAR_FLARES*. The HXR data before the year 2000 have been taken from *Burst and Transient Source Experiment (BATSE)* solar flare catalog. The URL address to collect these data is as follows *ftp://umbra.nascom.nasa.gov/pub/batse* and after year 2000 the data have been collected from *RESSI (Ramaty High-Energy Stectroscopic Imager)* data center (*http://hesperia.gsfc.nasa.gov/hessidata/dbase/hessi_flare_list.tex*). The analysis was carried out for the period from 01 October 2003 to 30 November 2003. During this period the occurrence of 401 SXR and 1920 HXR events are reported. All SXR and HXR events that overlapped in time with any other SXR and HXR events are excluded. To be identified as corresponding events, we have taken the condition that the start time difference between a SXR and a HXR event does not exceed 10 minutes (Veronig et al., 2002b). Applying these criterions we obtained 236 events that were observed in both SXR and HXR emissions.
In Figure 1, the monthly number of Hα, SXR flares and monthly mean sunspot numbers (SN) from 1996 to 2007 (SC 23) are plotted. This figure also represents the plot of monthly HXR flare numbers in the above mentioned period. It is clear from Figure 1 that Hα, SXR, SN and HXR show similar variation during the period of investigation.

## 3. STATISTICAL METHODS

To know the measure of dispersion we have applied the median absolute deviation, which can be calculated as

$$\bar{D} = Median\left\{\left|x_i - \bar{x}\right|\right\}, \quad (1)$$

where $x_i$ denote the data values and $\bar{x}$ is the median of the $x_i$. We have also used 95 % confidence intervals, which gives the probability for frequent use of the applied procedure. A conservative 95 % confidence interval for the median is given by a rule of thumb as $\bar{x} \pm c_{95}$ (Veronig et al., 2002a) with

$$c_{95} = \frac{1.58\ (Q_3 - Q_1)}{\sqrt{n}}, \quad (2)$$

$Q_1$ and $Q_3$ denote the first and the third quartile, respectively, n being the total number of data values. The first quartile $Q_1$, or $25^{th}$ percentile, of a distribution is given by the value which has 25 % of values below it; the third quartile $Q_3$, or the $75^{th}$ percentile, is given by the value with 75 % of values below it. The median $\bar{x}$ is identical to the second quartile $Q_2$, or the $50^{th}$ percentile. We have also calculated $90^{th}$ percentile $P_{90}$, which states that only 10 % of the events have a value larger than $P_{90}$. To study a characterization of the degree of asymmetry of a distribution around its mean we have also calculated the skewness of the distribution.
In order to characterize the properties of the rise and the decay time of a flare event, we computed the event asymmetry index $A_{ev}$ defined as

$$A_{ev} = \frac{t_{decay} - t_{rise}}{t_{decay} + t_{rise}}, \quad (3)$$

where $t_{rise}$ is the rise time, and $t_{decay}$ the decay time of events. The event asymmetry index is a dimension-less quantity. A value close to zero states that the rise and the decay times are roughly equal. And if $A_{ev} > 0$ the decay phase is longer then the rising phase else the rising phase will be longer than the decay phase (Temmer et al. 2001).
To investigate the NE we have used criteria i.e., the time difference ($\Delta t$) between the maximum of SXR emission and the end of the HXR emission should occur at the same time. For each event we determined the difference of the peak time of the SXR emission, $t_{SXR}$ and the end time of HXR emission, $t_{HXR}$

$$\Delta t = t_{SXR} - t_{HXR} \qquad (4)$$

A positive value indicates that the maximum of the SXR emission occurs after the end of the HXR emission, while a negative value indicates the occurrence of maximum of SXR and HXR in the reverse order (Veronig et al., 2002b).

## 4. ANALYSIS AND RESULTS

### 4.1 TEMPORAL PROPERTIES OF Hα FLARES

We statistically analysed temporal aspect of Hα flares, i.e. the duration, rise and decay times. Figure 2 shows the distributions of the duration, rise and decay times considering the total of Hα flares events. Each histogram reveals same type of variation. All distributions reveal a pronounced positive skewness. Solid line indicates the median value of the respective distribution. The dashed line indicates the $90^{th}$ percentile i.e., $P_{90}$ which states that only 10 % of the events have a value larger than $P_{90}$. In Table 5 we list various statistical measures characterizing the distributions of temporal parameters i.e., the arithmetic mean, the median, the mode and $P_{90}$ of the Hα flares. In Table 6, list of various statistical measures characterizing the distributions of temporal parameters, the arithmetic mean, the median, the mode for different importance classes are presented. In Table 8, same temporal parameters are listed with regard to the different brightness classes. Table 7 and 9 summarize the median values of the duration, rise time and decay time with respect to different importance and brightness classes respectively. In addition, the 95 % confidence interval, absolute median deviation and $90^{th}$ percentile of the duration, rise and decay times for the different importance and brightness classes are given.

**Table 5.** Mean, median, mode and $90^{th}$ percentile values of duration, rise, and decay times of the total number of Hα flares.

| Stat. measure | Duration (min) | Rise time (min) | Decay time (min) |
|---|---|---|---|
| Mean | 26.42 | 6.26 | 20.16 |
| Median | 12.00 | 2.00 | 9.00 |
| Mode | 8.00 | 1.00 | 7.00 |
| $P_{90}$ | 60.30 | 16.00 | 46.00 |

**Table 6.** Mean, median and mode values of duration, rise, decay times of the total number of different Hα flare importance classes (S, Imp. 1, Imp.>1).

| Class | Stat. measure | Duration (min) | Rise time (min) | Decay time (min) |
|---|---|---|---|---|
| S | Mean | 17.84 | 4.59 | 13.25 |
| | Median | 10.00 | 2.00 | 8.00 |
| | Mode | 7.00 | 1.00 | 7.00 |
| 1 | Mean | 48.50 | 8.65 | 39.85 |
| | Median | 43.00 | 6.00 | 34.00 |
| | Mode | 43.00 | 4.00 | 8.00 |
| >1 | Mean | 117.18 | 27.82 | 89.35 |
| | Median | 106.00 | 13.00 | 66.00 |
| | Mode | 170.00 | 13.00 | 66.00 |

**Table 7.** Median values with 95 % confidence interval, $\bar{x} \pm c_{95}$, absolute median deviation $\bar{D}$ and $90^{th}$ percentile of the duration, rise and decay times for the different importance classes (S, 1, >1) and the total number of flares (T). All values are given in minutes.

| Class | Duration | | | Rise time | | | Decay time | | |
|---|---|---|---|---|---|---|---|---|---|
| | $\bar{x} \pm c_{95}$ | $\bar{D}$ | $P_{90}$ | $\bar{x} \pm c_{95}$ | $\bar{D}$ | $P_{90}$ | $\bar{x} \pm c_{95}$ | $\bar{D}$ | $P_{90}$ |
| S | 10.00 ± 1.26 | 5.0 | 39.0 | 2.00 ± 0.39 | 2.0 | 12.0 | 8.00 ± 0.97 | 4.0 | 28.0 |
| 1 | 43.00 ± 9.28 | 18.0 | 94.0 | 6.00 ± 3.05 | 5.0 | 19.7 | 34.00 ± 7.79 | 16.5 | 80.2 |
| >1 | 106.00 ± 42.92 | 53.0 | 195.6 | 13.00 ± 13.03 | 12.0 | 68.6 | 66.00 ± 31.81 | 36.0 | 153.4 |
| T | 12.00 ± 1.93 | 7.0 | 60.3 | 2.00 ± 0.53 | 2.0 | 16.0 | 9.00 ± 1.42 | 5.0 | 46.0 |

**Table 8.** Mean, median and mode values of duration, rise, and decay times of the total number of different Hα flare brightness classes (f, n, b).

| Class | Stat. measure | Duration (min) | Rise time (min) | Decay time (min) |
|---|---|---|---|---|
| f | Mean | 19.70 | 4.91 | 14.80 |
| | Median | 11.00 | 2.00 | 8.00 |
| | Mode | 8.00 | 1.00 | 7.00 |
| n | Mean | 65.60 | 11.30 | 51.70 |
| | Median | 48.00 | 6.00 | 41.00 |
| | Mode | 11.00 | 4.00 | 9.00 |
| b | Mean | 127.20 | 27.50 | 99.70 |
| | Median | 113.50 | 17.50 | 98.50 |
| | Mode | --- | 25.00 | --- |

**Table 9.** Median values with 95 % confidence interval, $\bar{x} \pm c_{95}$, absolute median deviation $\bar{D}$ and $90^{th}$ percentile of the duration, rise and decay times for the different importance classes (f, n, b) and the total number of flares (T). All values are given in minutes.

| Class | Duration | | | Rise time | | | Decay time | | |
|---|---|---|---|---|---|---|---|---|---|
| | $\bar{x} \pm c_{95}$ | $\bar{D}$ | $P_{90}$ | $\bar{x} \pm c_{95}$ | $\bar{D}$ | $P_{90}$ | $\bar{x} \pm c_{95}$ | $\bar{D}$ | $P_{90}$ |
| f | 11.00 ± 1.50 | 6.0 | 44.2 | 2.00 ± 0.47 | 2.0 | 12.0 | 8.00 ± 1.03 | 4.0 | 35.0 |
| n | 48.00 ± 21.74 | 28.0 | 167.8 | 6.00 ± 4.12 | 4.0 | 26.6 | 41.00 ± 17.63 | 24.0 | 116.2 |
| b | 113.50 ± 51.21 | 56.0 | 231.1 | 17.50 ± 9.62 | 11.0 | 58.3 | 99.70 ± 46.97 | 49.5 | 159.0 |
| T | 12.00 ± 1.95 | 7.0 | 60.3 | 2.00 ± 0.53 | 2.0 | 16.0 | 9.00 ± 1.41 | 5.0 | 46.0 |

Figure 3 shows the distributions of the durations separately for different classes of SXR flares. All distributions reveal a pronounced positive skewness. In this figure, for the sake of clearness, we are using the cut off values of the duration for different classes. It can also be seen from this figure that with increasing flare class the center of distribution shift to a larger value. The differences from one class to the other are larger than the 95 % confidence limit, indicating the statistical significance of the effect for importance as well as brightness classes. From Tables 6 and 8 it is clear that the mean of the duration, rise and decay time of importance as well as brightness classes increases with the flare class respectively.

### 4.2 TEMPORAL PROPERTIES OF SXR FLARES

We statistically analysed temporal aspect of SXR flares, i.e. the duration, rise and decay times. Figure 4 shows the distributions of the duration, rise and decay times considering the total of SXR events. The solid lines indicate the median values of the respective distributions. The dashed line indicates the $90^{th}$ percentile i.e., $P_{90}$. Each histogram reveals similar variation. In Table 10 we list various statistical measures characterizing the distributions of the temporal parameters, the arithmetic mean, the median, the mode and the $90^{th}$ percentile i.e. $P_{90}$ of the SXR flares. In Table 11 lists of various statistical measures characterizing the distributions of temporal parameters, the arithmetic mean, the median, and the mode for different SXR flares classes are presented. Table 12 summarizes the median values of the duration, rise time and decay time with respect to different importance classes. Additionally, the 95 % confidence intervals, absolute median deviations and $90^{th}$ percentile of the duration, rise and decay times for the different SXR classes are also given.

**Table 10.** Mean, median, mode and the $90^{th}$ percentile values of duration, rise, and decay times of the total number of SXR flares.

| Stat. measure | Duration (min) | Rise time (min) | Decay time (min) |
|---|---|---|---|
| Mean | 22.68 | 11.42 | 11.26 |
| Median | 13.00 | 6.00 | 6.00 |
| Mode | 8.00 | 4.00 | 4.00 |
| $P_{90}$ | 43.00 | 24.00 | 22.00 |

**Table 11.** Mean, median and mode values of duration, rise, and decay times of the total number of different SXR flares classes (B, C, M, X).

| Class | Stat. measure | Duration (min) | Rise time (min) | Decay time (min) |
|---|---|---|---|---|
| B | Mean | 20.34 | 9.42 | 10.93 |
| | Median | 11.50 | 5.00 | 5.00 |
| | Mode | 6.00 | 3.00 | 4.00 |
| C | Mean | 18.38 | 9.40 | 8.99 |
| | Median | 12.00 | 6.00 | 6.00 |
| | Mode | 8.00 | 4.00 | 4.00 |
| M | Mean | 37.23 | 18.90 | 18.33 |
| | Median | 22.00 | 11.00 | 11.00 |
| | Mode | 11.00 | 6.00 | 4.00 |
| X | Mean | 55.55 | 30.27 | 21.27 |
| | Median | 36.00 | 21.00 | 14.00 |
| | Mode | 36.00 | 21.00 | 14.00 |

**Table 12.** Median values with 95 % confidence interval, $\bar{x} \pm c_{95}$, absolute median deviation $\bar{D}$, and $95^{th}$ percentile of the duration, rise and decay times for the different SXR classes (B, C, M, X) and the total number of flares (T). All values are given in minutes.

| Class | Duration | | | Rise time | | | Decay time | | |
|---|---|---|---|---|---|---|---|---|---|
| | $\bar{x} \pm c_{95}$ | $\bar{D}$ | $P_{90}$ | $\bar{x} \pm c_{95}$ | $\bar{D}$ | $P_{90}$ | $\bar{x} \pm c_{95}$ | $\bar{D}$ | $P_{90}$ |
| B | 11.50 ± 1.69 | 4.5 | 31.0 | 5.00 ± 0.46 | 2.0 | 13.5 | 5.00 ± 1.30 | 2.0 | 18.5 |
| C | 12.00 ± 1.38 | 5.0 | 34.6 | 6.00 ± 0.63 | 2.0 | 17.8 | 6.00 ± 0.63 | 3.0 | 17.8 |
| M | 22.00 ± 6.27 | 12.0 | 73.0 | 11.00 ± 2.83 | 5.0 | 37.0 | 11.00 ± 3.03 | 7.0 | 33.0 |
| X | 36.00 ± 15.48 | 6.0 | 96.0 | 21.00 ± 11.67 | 7.0 | 58.0 | 14.00 ± 2.89 | 2.0 | 39.0 |
| T | 13.00 ± 1.26 | 6.0 | 43.0 | 6.00 ± 0.55 | 2.0 | 24.0 | 6.00 ± 0.63 | 3.0 | 22.0 |

Figure 5 shows the distributions of the duration separately for different classes of SXR flares. All distributions reveal a pronounced positive skewness. In this figure for the sake of clearness, we are using the cut off values of the duration for different classes. It can also be seen from this figure that with increasing flare class the center of distribution moves to a larger value. The differences from one class to the other are larger than the 95 % confidence limit, indicating the statistical significance of the effect. However due to the poor statistics of X class flares; the respective 95 % confidence limits are somewhat larger. From Table 11 it can be seen that the mean of the duration, rise and decay time increases with the flare class (from B to X). From Tables 10, 11 and 12 it follows that the median values of the rise and decay times are quite similar, for the overall number of flares as well as for different flare classes.

### 4.3 TEMPORAL PROPERTIES OF HXR FLARES

We statistically analysed temporal aspect of HXR flares, i.e. the duration, rise and decay times. In Figure 7, we have presented the distribution of duration for HXR during the same period under consideration. The variation of this plot is the same as that of Hα and SXR. The solid line indicates the median value of the distributions. The dashed line indicates the $90^{th}$ percentile i.e., $P_{90}$. In Table 13 we list various statistical measures characterizing the distributions of temporal parameters, the arithmetic mean, the median, the mode and the $90^{th}$ percentile i.e., $P_{90}$ of the HXR flares. Table 14 summarizes the median values of the duration, rise time and decay time of HXR events. The 95 % confidence intervals, absolute median deviations and $90^{th}$ percentile of the duration, rise and decay times for total HXR flares are also given.

**Table 13.** Mean, median, mode and the $90^{th}$ percentile values of duration, rise, and decay times of the total number of SXR flares.

| Stat. measure | Duration (min) | Rise time (min) | Decay time (min) |
|---|---|---|---|
| Mean | 5.69 | 2.34 | 3.34 |
| Median | 1.53 | 0.63 | 0.57 |
| Mode | 0.33 | 0.17 | 0.17 |
| $P_{90}$ | 13.33 | 4.90 | 7.70 |

**Table 14.** Median values with 95 % confidence interval, $\bar{x} \pm c_{95}$, absolute median deviation $\bar{D}$, and $P_{90}$ percentile of the duration, rise and decay times for HXR flares. All values are given in minutes.

| Duration | | | Rise time | | | Decay time | | |
|---|---|---|---|---|---|---|---|---|
| $\bar{x} \pm c_{95}$ | $\bar{D}$ | $P_{90}$ | $\bar{x} \pm c_{95}$ | $\bar{D}$ | $P_{90}$ | $\bar{x} \pm c_{95}$ | $\bar{D}$ | $P_{90}$ |
| $1.53 \pm 0.17$ | 1.2 | 13.3 | $0.63 \pm 0.06$ | 0.5 | 4.91 | $0.57 \pm 0.09$ | 0.5 | 7.7 |

### 4.4 EVENT ASYMMETRY

We have calculated the event asymmetry index using Equation 3. Hα flare does not reveal a symmetrical behavior regarding its temporal evolution, i.e. the rise time is not the same as the decay time whereas SXR and HXR flares show symmetric behavior. In Table 15 and Table 16 we list the median values of the event asymmetries as well as 95 % confidence intervals, the absolute median deviation and the $10^{th}$ percentile for the different classes in Hα and SXR respectively. Table 17 represents the same parameters for HXR flares. Figure 6 represents the distribution of the event asymmetries calculated for Hα, SXR and HXR flares during 2003 October–November. All distributions reveal a pronounced negative skewness, i.e., an accumulation at positive values, showing that for the majority of events the decay phase is significantly longer than the rising phase.

**Table 15.** Median values of event asymmetries for different importance classes with 95 % confidence intervals. Furthermore, the absolute median deviations and the $10^{th}$ percentiles are listed.

| Class | Event asymmetries | | |
|---|---|---|---|
| | $\bar{x} \pm c_{95}$ | $P_{10}$ | $\bar{D}$ |
| S | $0.600 \pm 0.062$ | -0.012 | 0.0 |
| 1 | $0.708 \pm 0.086$ | 0.232 | $3.1\times10^{-7}$ |
| >1 | $0.644 \pm 0.163$ | 0.164 | $4.4\times10^{-7}$ |
| T | $0.610 \pm 0.053$ | 0.000 | $-1.9\times10^{-7}$ |

**Table 16.** Median values of event asymmetries for the different SXR classes with 95% confidence intervals. Also listed are, the absolute median deviations and the $10^{th}$ percentiles.

| Class | Event asymmetries | | |
|---|---|---|---|
| | $\bar{x} \pm c_{95}$ | $P_{10}$ | $\bar{D}$ |
| B | $0.000 \pm 0.643$ | -0.333 | 0.0 |
| C | $0.000 \pm 0.035$ | -0.667 | 0.0 |
| M | $-0.091 \pm 0.064$ | -0.375 | $9.09\times10^{-7}$ |
| X | $-0.167 \pm 0.076$ | -0.222 | $3.33\times10^{-6}$ |
| T | $0.000 \pm 0.027$ | 0.388 | 0.00 |

**Table 17.** Median values of event asymmetries for HXR flares with 95 % confidence intervals. Furthermore, the absolute median deviations and the $10^{th}$ percentiles are also listed.

| Event asymmetries | | |
|---|---|---|
| $\bar{x} \pm c_{95}$ | $P_{10}$ | $\bar{D}$ |
| $0.154 \pm 0.029$ | -0.813 | 0.38 |

For the total number of Hα events we obtain a median event asymmetry of ≈ 0.61, which implies that for about 50 % of the events the decay phase is more than 4.13 times as long as the rising phase and a value of $P_{10}$ ≈ 0.0 means that only about 10 % of all flares have a shorter decay than rise time. For SXR, the value of median event asymmetry is ≈ 0.0, which means that for about 50 % of the events the decay phase is equal to the rising phase and for HXR this value is ≈ 0.15 which implies that for about 50 % of the HXR events the decay phase is more than 1.35 times longer than the rising phase. For SXR and HXR the values of $P_{10}$ are 0.39 and -0.81 respectively which means that for only about 10 % of all flares, the decay phase is less than 2.28 and 0.10 times longer than the rising phase respectively.

### 4.5 NEUPERT EFFECT

We have calculated the value of the time difference ($\Delta t$) using Equation 4. Figure 8 shows histogram of time difference between the SXR maximum and HXR end times ($\Delta t$) verses number of flares. Positive values indicate that the maximum of SXR emission occurs after the end of the HXR emission, and negative values indicate the reverse phenomena. Our analysis shows that there are more events for which SXR maximum takes place after the HXR end (113 (47.89 %)) than vice versa (72 (30.51 %)). Our analysis also shows that out of 236 events, 48 events satisfy the NE i.e., the difference of SXR peck time and the HXR end time is less than 1 min and more than half of the events do not satisfy this effect. The outcome suggests that during this period 20.34 per cent events reveal a timing behavior that is consistent with NE, but there exists also a significant fraction of events that are incompatible with NE. However, there is also a significant fraction 62 (26.27 %) of the flares, which show strong deviations from $\Delta t = 0$.

## 5. DISSCUSSIONS AND CONCLUSIONS

We have analysed the temporal aspects of flares in different wavelengths (Hα, SXR and HXR) and the results obtained are as follows:

- On an average, the duration, rise, and decay times of Hα flares increase with increasing importance classes (S, 1, >1). The increase is more pronounced for the duration and the decay time of about 9-10 times between subflares and importance >1. Similar variation holds for brightness classes also.
- On the other hand the duration, rise, and decay times of SXR flares does not increase with increasing importance classes (B, C, M, X). The values first decrease from B to C class and than increase up to X class. The increase is more pronounced for the duration and the rise time is of about 3-5 times between B and X classes.
- The event asymmetries, which characterize the proportion of the decay to the rise time of a flare, are predominantly positive for Hα while for SXR and HXR these are distributed on both sides of the zero asymmetry.
- The NE analysis of our study during October-November 2003 shows that out of 236 events, only 48 events satisfy the NE. The distribution of the difference of the SXR peak times and HXR end times is strongly peaked at $\Delta t = -1$.

The facts that on an average the Hα flare duration increases with the importance class and the rise times are shorter than the decay times are reported in a number of previous papers (Waldmeier 1938; Ellison 1949; Smith 1962; Antalová 1985). Recently Temmer et al. (2001) analysed the Hα flare evolution parameter i.e. duration, rise time, decay times and found that these parameter increases with increasing importance class (S, 1, >1) with the result that increase is more pronounced for the decay times than for the rise times. Our study shows that Hα flare parameter increases with increasing importance class but the increase is more pronounced for duration as well as decay times than for the rise times during the period of our investigation. These results suggest that, with respect to the temporal behavior, the cooling phase of the Hα flare is more strongly affected by the flare size than the phase of heating-up of the chromospheric plasma at the flare site. Temmer et al. (2001) also analysied event asymmetries and found these predominantly positive. Event asymmetry does not reveal a symmetrical behavior regarding their temporal evolution, i.e. the rise time is not the same as the decay time. Our study shows similar result regarding the event asymmetry.

Veronig el al (2002a) presented a statistical analysis of SXR flare evolution steps i.e. duration, rise time, and decay time and found that these steps increase with increasing importance classes (B, C, M, X). The increase is more pronounced for the rise phase (factor of 4 between B and X class) while increase in duration as well as decay phase is less pronounced (factor of 3 between B and X class). In our study rise time is more pronounced (factor of 4 between B and X class) and in duration as well as decay phase is less pronounced (factor of 3 between B and X class).

From Figure 2, 3, 4, 5, and 8, it is clear that all the distributions in Hα, SXR and HXR show same type of variation in duration, rise and decay time and the duration of their respective classes. From Figure 6 it can be clearly seen that the variation of EA for SXR and HXR is the same but different from Hα. This is because we have a larger number of Hα flares with zero rise time.

Veronig et al, (2002b) studied several aspects of the NE and their distribution study of the time difference ($\Delta t$) reveals a pronounced peak at ($\Delta t = 0$). They also investigate that about half of the events show a timing behavior which can be considered to be consistent with the expectations from the NE. In our analysis time difference peak lies at $\Delta t = -1$ and only 20.34 % of events chosen indicate NE whereas more than half of the event do not.

**ACKNOWLEDGEMENTS:**

Authors thank UGC, New Delhi for financial assistance under DSA (phase-III) program running in the Department of Physics, Kumaun University, Nainital. NCJ and NS are thankful to UGC, New Delhi for financial assistance under RFSMS (Research Fellowship in Science for meritorious students) scheme.

---------------

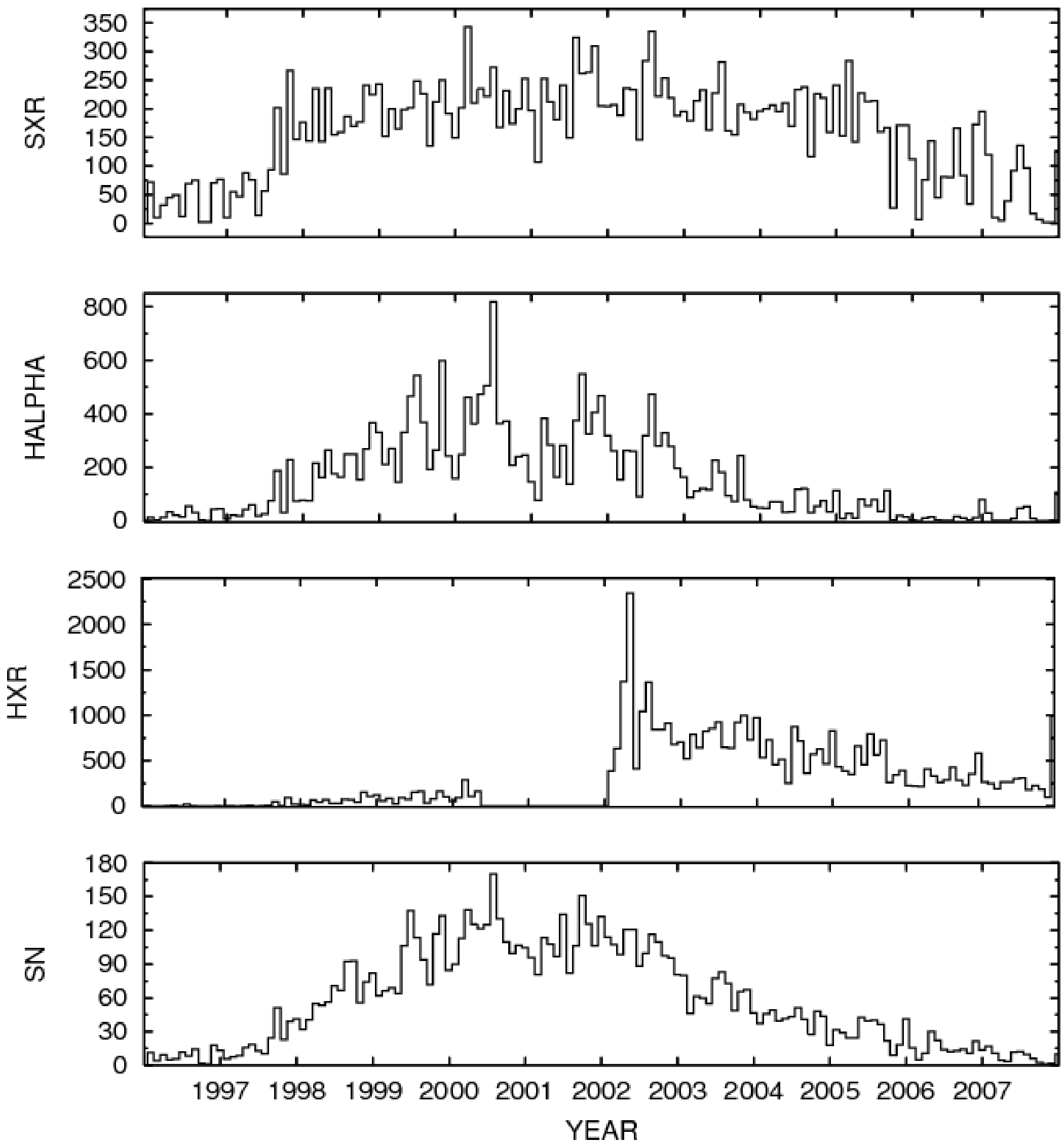

**Figure 1.** Monthly plot for flares of different types, SXR flares, Hα flares, HXR flares and monthly mean sunspot numbers (SN) (from top to bottom panel) from 1996-2007.

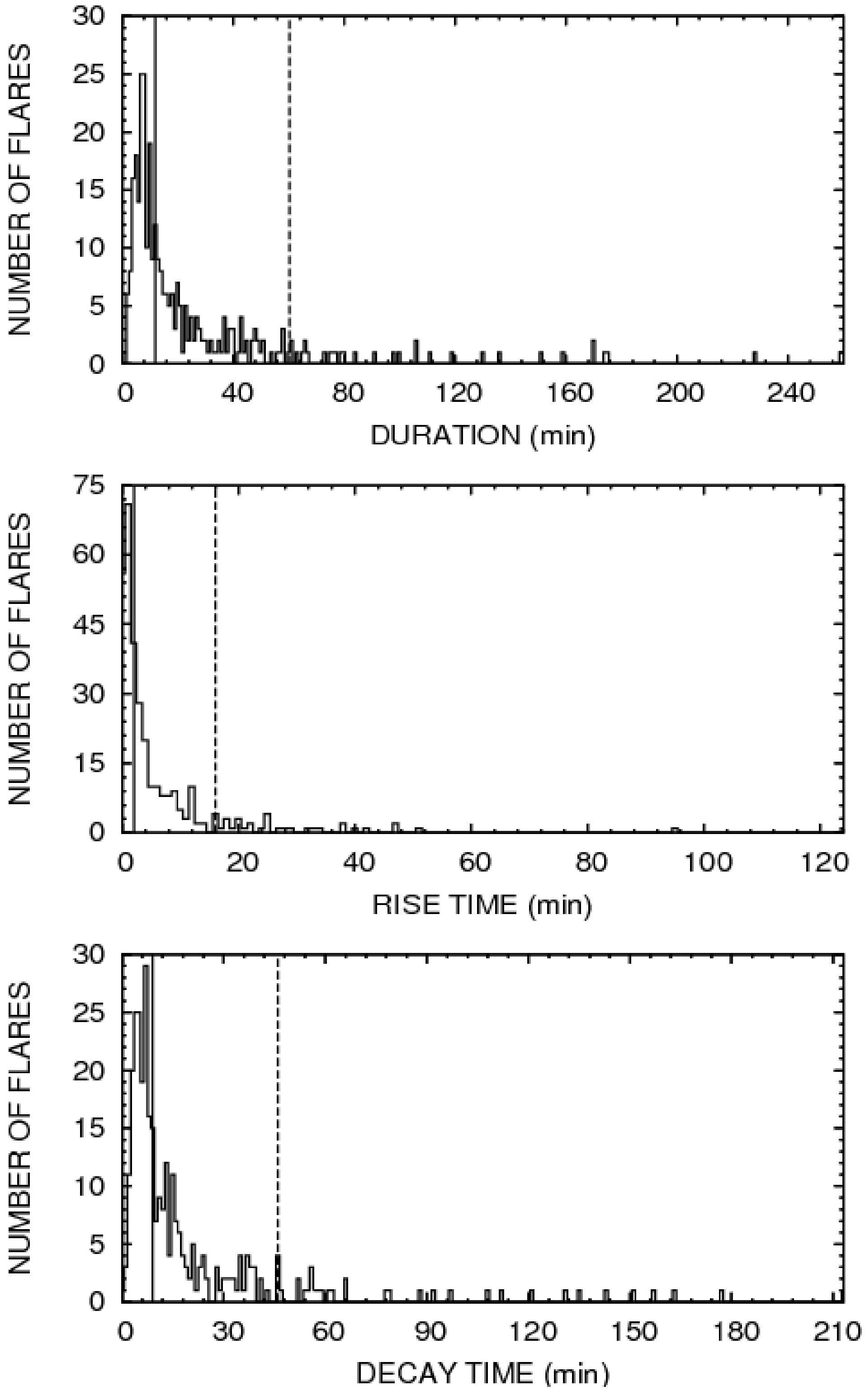

**Figure 2.** Histograms of duration, rise and decay times of Hα flares during October-November 2003. The solid line indicates the median value of the distribution, the dashed line the 90th percentile.

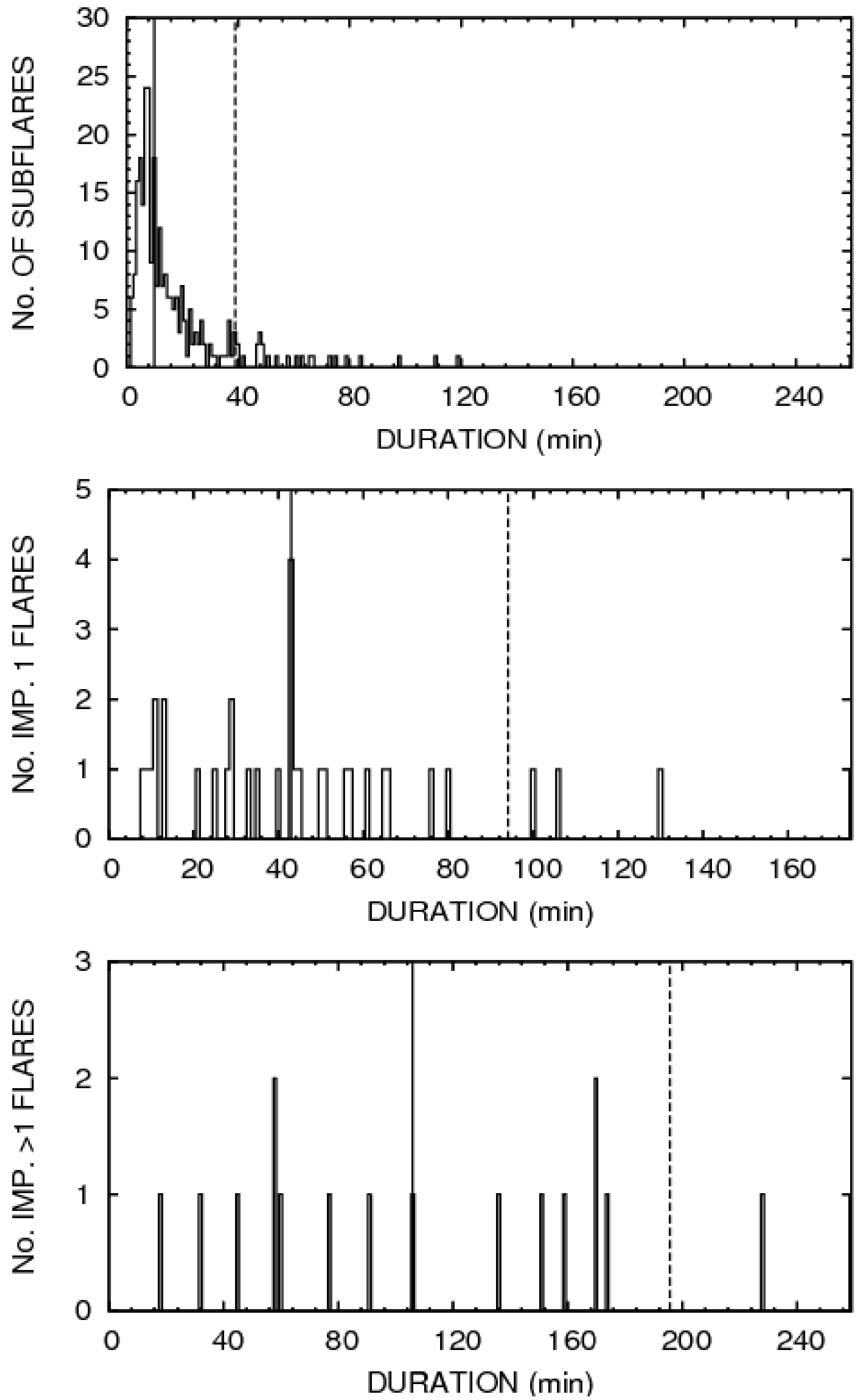

**Figure 3.** Histograms of duration for the different Hα flare classes during October-November 2003. The solid line indicates the median value of the distribution, the dashed line the 90$^{th}$ percentile.

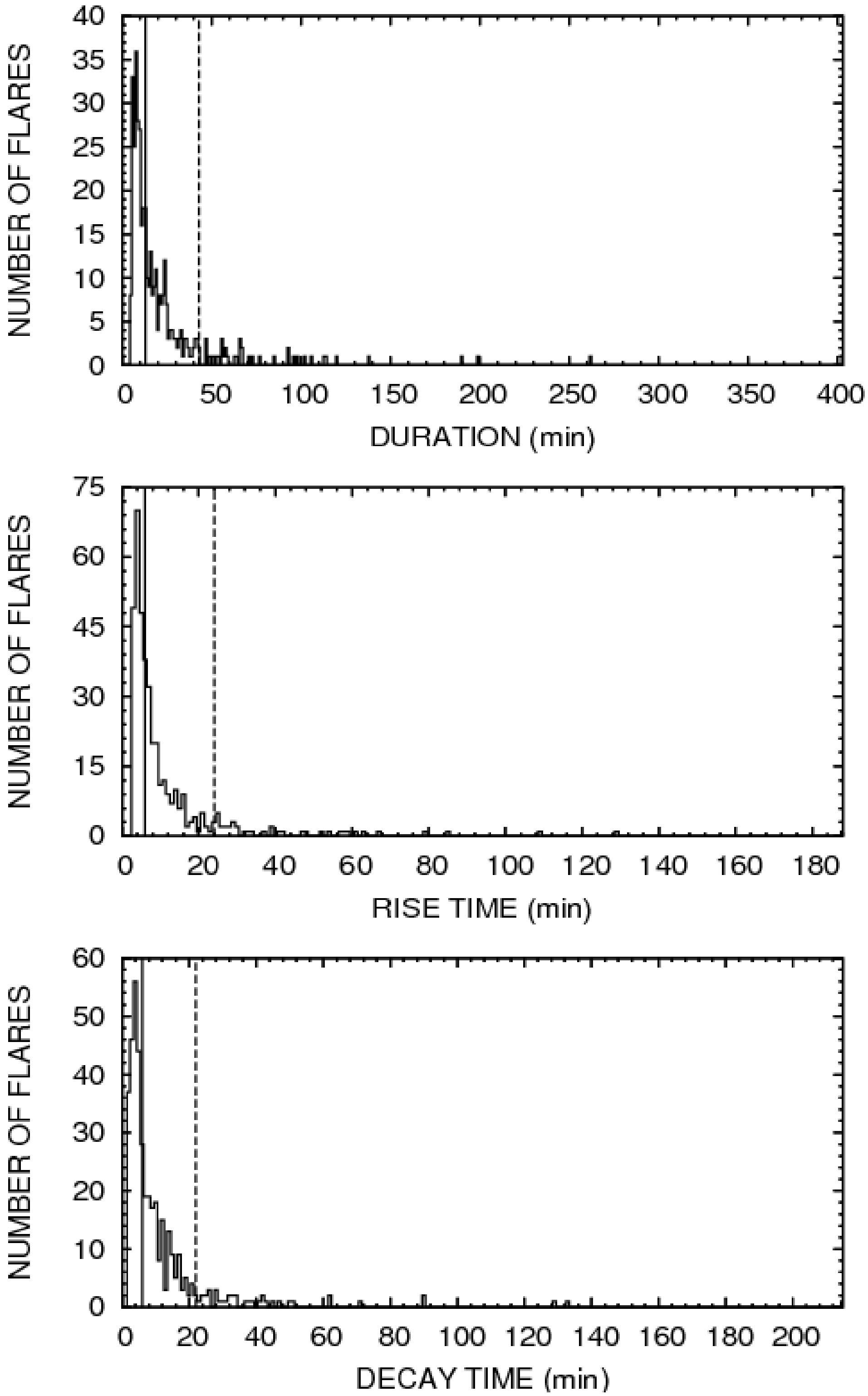

**Figure 4.** Same as Figure 2, but for SXR flares.

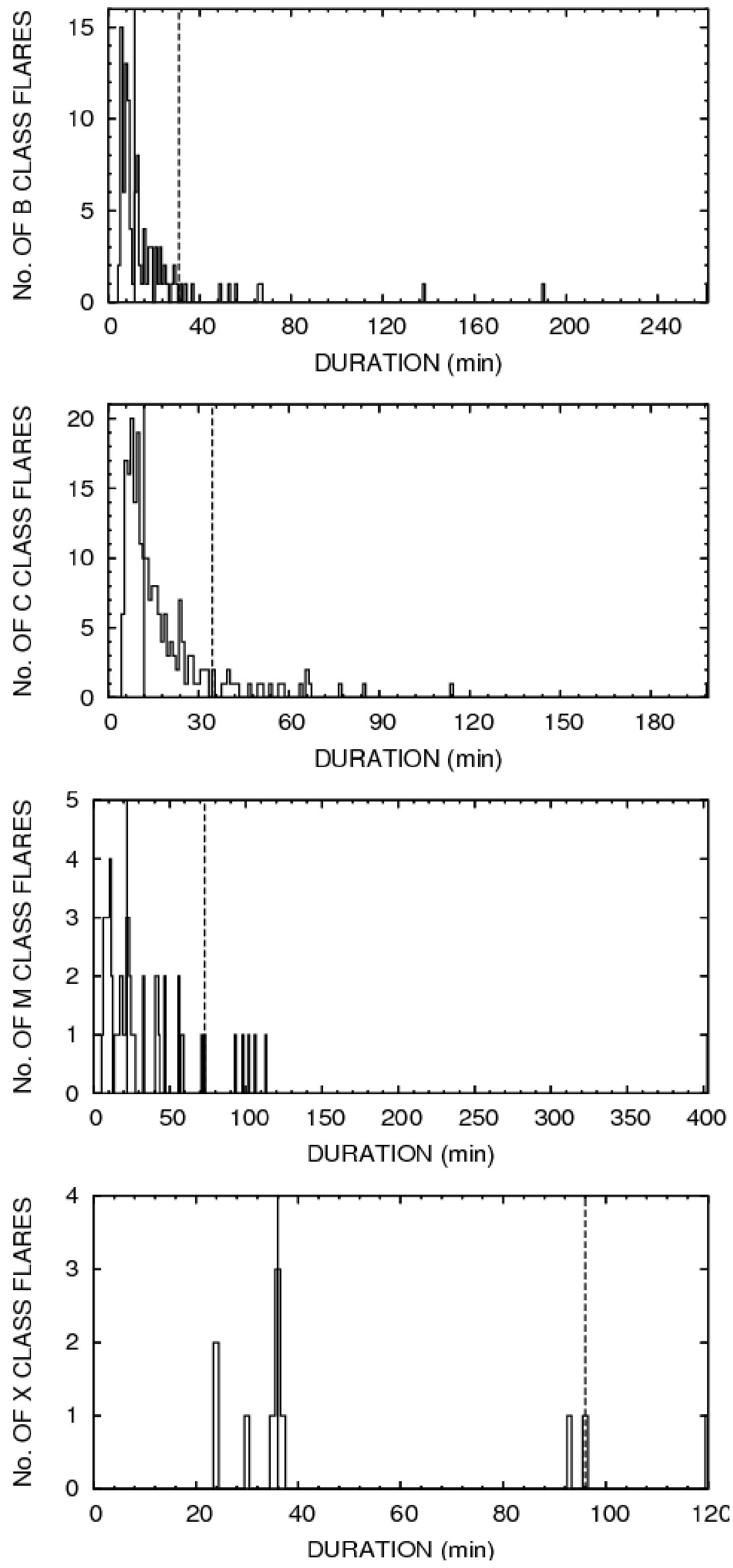

**Figure 5.** Same as Figure 3, but for SXR flares.

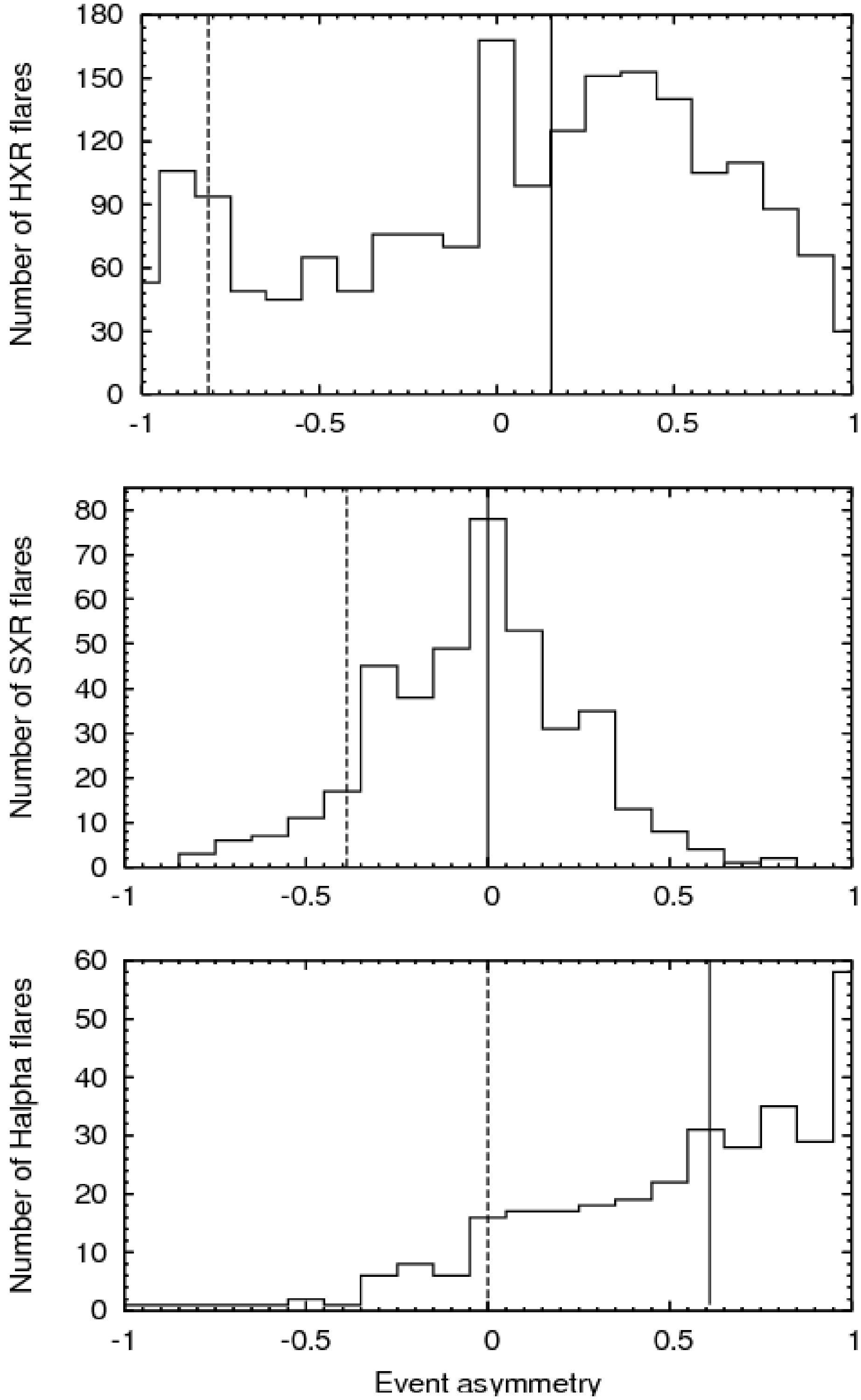

**Figure 6.** Histogram of the event asymmetry for HXR, SXR and Hα flares during October-November 2003 (from top to bottom panel). The solid line indicates the median of the distribution, the dashed line the $10^{th}$ percentile $P_{10}$, which indicates that only 10 per cent of the events have a value smaller than $P_{10}$.

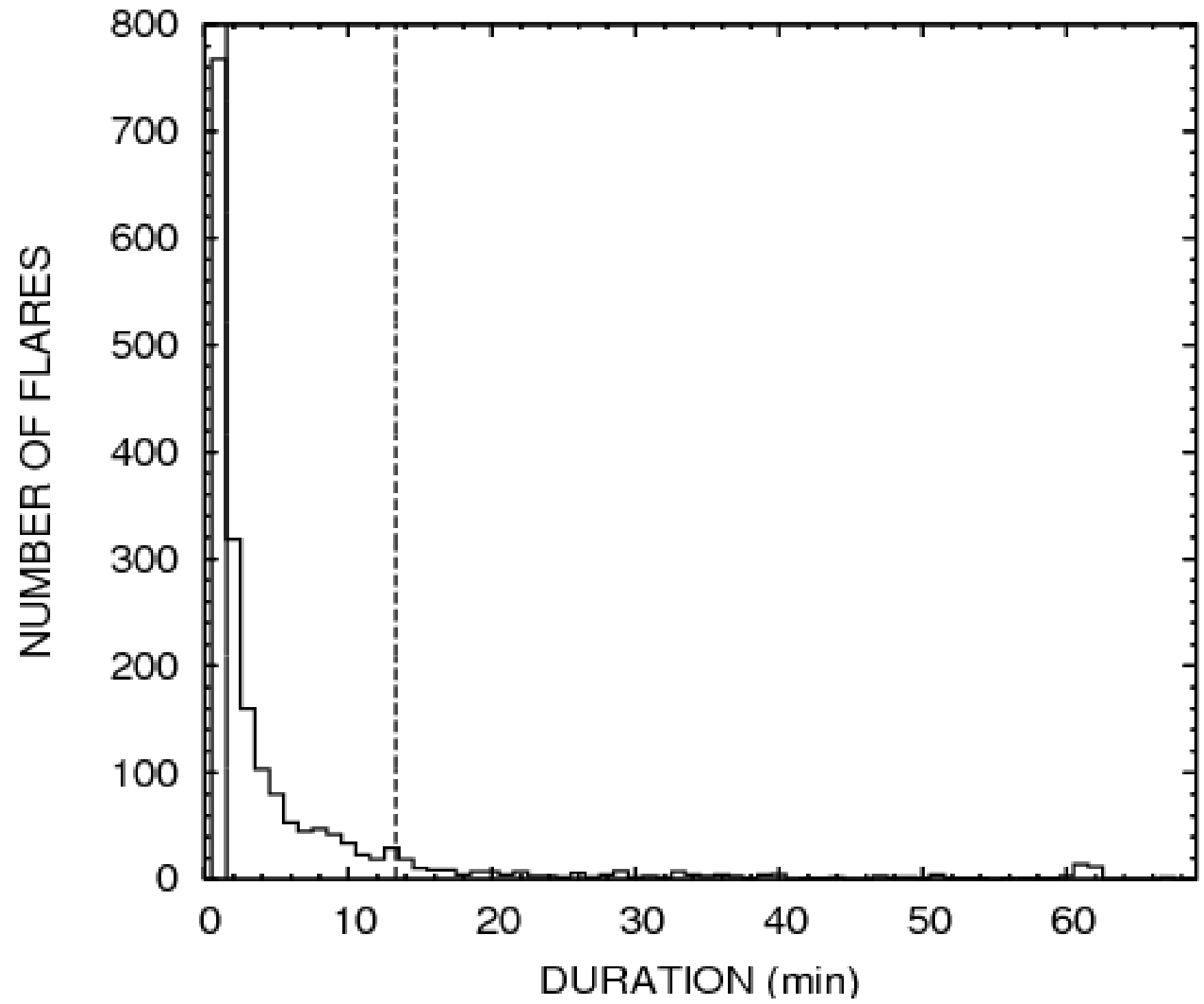

**Figure 7.** Histogram of duration for HXR flares during October-November 2003. The solid line indicates the median value of the distribution, the dashed line the 90th percentile.

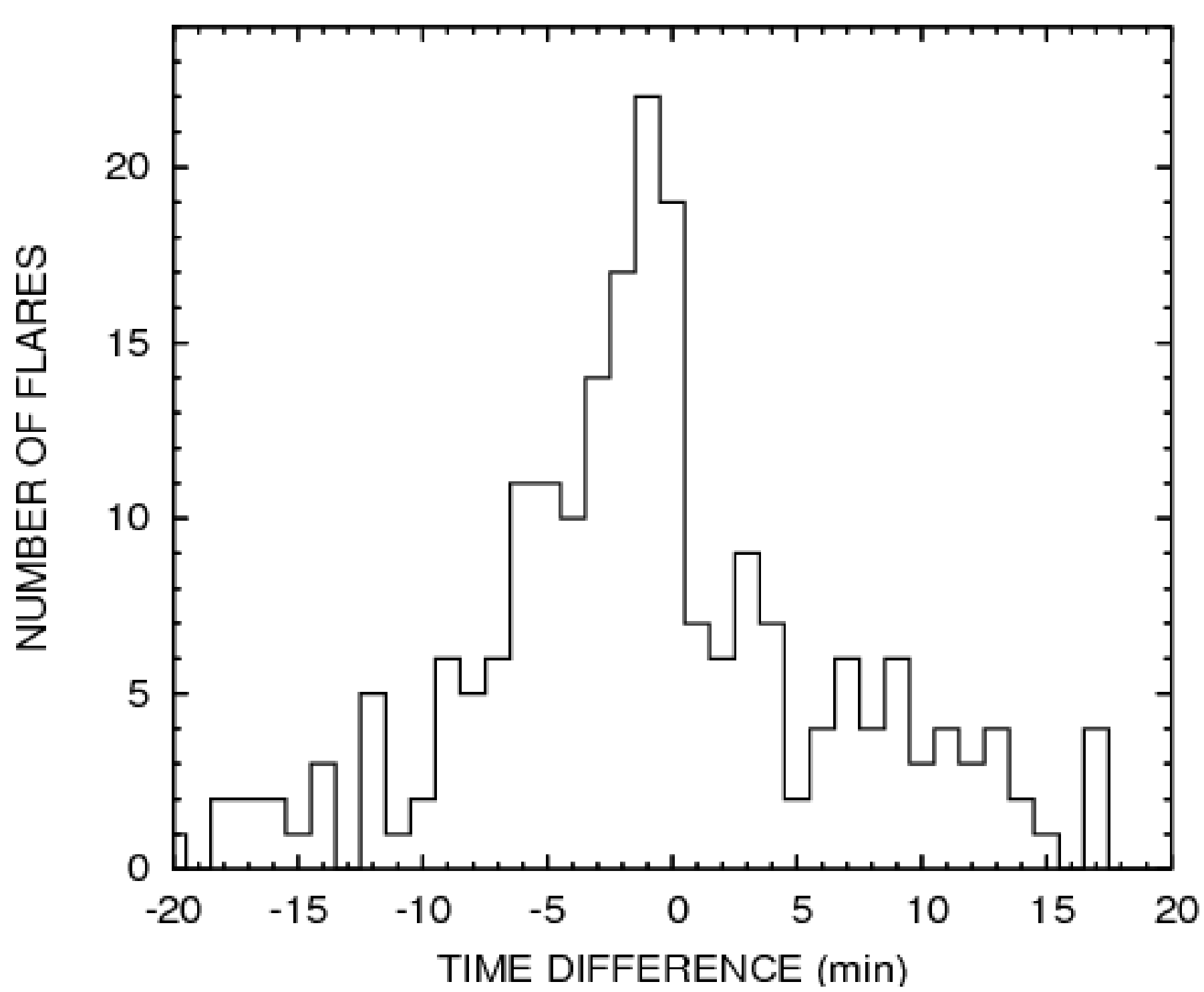

**Figure.8** Histogram of the difference of the SXR maximum and HXR end time.